# Southern Binary Galaxies
# I. A Sample of Isolated Pairs

D.S.L. Soares[1], R.E. de Souza[2], R.R. de Carvalho[2,3], and T.C. Couto da Silva[2]

1 Departamento de Física, UFMG, C.P. 702, Belo Horizonte, 30161-970, Brazil
2 Instituto Astronômico e Geofísico, USP, C.P. 9638, São Paulo, 01065-970, Brazil
3 Observatório Nacional/DAF, CNPq, C.P. 23002, Rio de Janeiro, 20921-400, Brazil

# Abstract

A catalogue of binary galaxies with 621 pairs has been determined by applying a surface density enhancement procedure to *The Surface Photometry Catalogue of the ESO-Uppsala Galaxies*. The method does not require any redshift information. An additional restriction, based on objective criteria that take into account the completeness of the source catalogue, led to a sample of 189 isolated pairs which are listed. We have obtained the optical luminosity function of binary galaxies in the catalogue, from which we estimate that the luminosity density of galaxies in binaries is ∼4% of that found for field galaxies. The general properties of our sample are similar to those from CPG and CMG.

# 1 Introduction

The clustering properties of the galaxy distribution in the Universe is one of the main topics studied by the observational cosmologists in the recent years. Clustering scale varies from "totally" isolated galaxies to clusters of clusters of galaxies. This interval is usually binned in different classes of multiplicity as a matter of simplicity, not necessarily implying any physical difference. In such hierarchy, pairs of galaxies represent the first step after the isolated galaxies with no clear cut between them. Their difference is only a matter of what kind of isolation criterion is used. However, even without a totally objective way to define an isolated pair of galaxies they can still be useful in shaping our understanding of the physics of galaxy interactions and formation mechanisms.

The historical step in the way of understanding how interacting galaxies evolve and have their internal and orbital structure changed, came with the Toomre and Toomre's (1972) investigation on the M51 and Antennae systems. Modelling of interacting binary systems has now been done by many authors (e.g., Barnes 1988, Borne 1988, Balcells, Borne and Hoessel 1989, Madejsky 1992, Keel and van Soest 1992). Close and wide binaries have been useful for the study of many aspects of galaxy morphology and its connection to galaxy formation and the Hubble sequence (Sulentic 1992, and references therein). Such systems are also useful for statistical analysis aiming the determination of galaxy mass (Karachentsev 1972, 1985, 1987, Turner 1976, Schweizer 1987, Soares 1989), and the study of the secular evolution of orbital parameters (e.g., Verner and Chernin 1987).

Given the relatively large number of available binary lists, one might question whether the determination of yet another double galaxy list is really necessary. In fact, there exists a wealthy number of galaxy pair objective catalogues, i.e., catalogues determined with automated and well-defined procedures. However, most of them covers the northern sky (e.g., Karachentsev 1972, CPG - Catalogue of Isolated Pairs of Galaxies), while the southern sky lacks of such an extensive survey. The available southern lists are either incomplete or based on subjective selection criteria (Arp and Madore 1985, Schweizer 1987, Zhenlong et et al. 1989, Rampazzo and Sulentic 1992).

In section 2 we make a detailed description of the selection procedures used to build our binary sample. Section 3 is devoted to the determination and discussion of the luminosity function of galaxies in our binary catalogue. Finally, section 4 contains a brief discussion of the general characteristics of the sample.

## 2 Sample Selection

The selection of the present binary galaxy sample was based on the galaxy-galaxy projected separations and apparent magnitudes. The source list used for this purpose was *The Surface Photometry Catalogue of the ESO-Uppsala Galaxies* (hereafter ESO-LV, Lauberts and Valentijn 1989). The ESO-LV has 15467 entries, most of them with total magnitudes in B passband. The catalog is probably complete until 14.5 mag, as will be discussed in the next section and covers a solid angle of 2.01 sterradians. It is important to keep in mind that such incompleteness may introduce a bias in our binary sample, i.e., we may lose pairs for which the secondary galaxy is much fainter than the limiting magnitude.

The likelihood of physical association for the candidate pairs was estimated following the prescriptions used in the construction of the *Catalog of Multiple Galaxies* (CMG, van Albada and Soares 1993, and Soares 1989; see also van Moorsel 1982, van Moorsel 1987, Oosterloo 1988). A summary of the main points of such method is presented in the Appendix. For a more detailed discussion the reader is refered to Soares (1989).

In what follows we show how the CMG method was applied to the ESO-LV. The method consists of investigating the frequency distribution of projected separations taking into account the apparent magnitude of each galaxy in the pair. This is done in the following way. Each galaxy is classified according to its apparent magnitude, and put in 0.5-mag bins, whose centers range from $B_{T,1} = 12.0$ up to $B_{T,6} = 16.5$. They are numbered by $J = 1, 2, ..., 10$. Thus, the interval $J = n$ has all galaxies with magnitudes from $B_{T,n} - 0.25$ to $B_{T,n} + 0.25$. Now, a pair of galaxies is characterized by a given $\Delta J$, defined as the difference between J(secondary) − J(primary). The distributions of separations for every $\Delta J$, from $\Delta J = 0$ to $\Delta J = 9$, were investigated individually. We only considered pairs for which the primary galaxy was brighter than 14.5, minimizing the effect of incompleteness of the source catalog in our final binary sample. However, the secondary was selected regardless its magnitude. Galaxies with no magnitude listed by the ESO-LV were included in the last bin ($J = 10$) and galaxies brighter than the 11.75 have been included in the first bin ($J = 1$) in order to prevent the undersampling at the brighter end.

The distribution of first-neighbour projected separations is supposed to be Poissonian, in the absence of physical association. Any deviation, in excess over that, reflects small-scale gravitational clustering. The likelihood of physical association of a given pair is defined from the amount of deviation of the actual separation distribution over the random (Poissonian) distribution. Let $p$ be such a *probability*, given by

$$p = 1 - \frac{p_1(x)}{p_\circ(x)} , \qquad (1)$$

where $x$ is the normalized projected separation (see below), $p_1(x)$ is the theoretical Poissonian distribution of $x$ (fitted to large values of $x$). The observed values of $x$ are obtained by dividing $\theta_1$, the first-neighbour separation, by the Poissonian expected mean value of $\theta_1$, which in turn depends on the surface density of galaxies in the vicinity of the pair (n). Such a normalization of $\theta_1$, implies that the frequency distribution of $x$ is *independent* of the local surface density of galaxies, making it possible to treat galaxies from different clustering environments into one distribution. Following the CMG recipe, we have used $x$ as given by eq. (A5),

$$x = \theta_1 2\sqrt{n} = 3.52 \frac{\theta_1}{\theta_{10}} , \qquad (2)$$

in order to test for physical association. The actual distributions $p_\circ(x)$ were obtained for pairs belonging to $\Delta J = 0, 1, ..., 5$. We have adopted also a crude grouping of the morphological types in 3 broad classes, namely, early-type galaxies ($-5 < T < 0$), spirals ($1 < T < 6$) and late spirals and irregulars ($T > 6$). The excess of $p_\circ(x)$ over the Poissonian distribution was fitted to a linear approximation for each morphological type of the primary. The solid lines shown in Figure 1 are given by

$$p(x,T) = 1 - \frac{x}{2x_\circ(T)} \quad . \tag{3}$$

We can see from Figure 1 that the behaviour of $p(x,T)$ is independent of the magnitude bin of the secondary, contrary to what was found for the CMG, where it is noted a substantial correlation of $p(x,T)$ with $\Delta J$ (Soares 1989). Moreover, the relation is steeper for early type galaxies, since these objects are more clustered than the late type ones; such a behaviour was also noticed in the determination of the CMG.

This parameter, $p(x,T)$, was used as our primary indicator of association. If for a given bin of magnitude, $p(x,T)$ is larger than a critical value ($p_c = 0.5$) then the pair is selected as a probable candidate to form a binary system. In order to eliminate binaries in dense groups and clusters we have adopted a local density cutoff of 5 galaxies per square degree. This density measurement was obtained from the distance to the tenth nearest neighbour (see Appendix, or Soares 1989, for a complete discussion). Imposing all these constraints we have obtained a crude sample containing 621 binaries. Further examination has shown that 70 binary pairs from this first list were actually members of multiple systems. This inspection was done looking for different pairs sharing the same galaxy. Excluding these obviously false pairs we got a purged list of 551 pairs.

An important effect jeopardizing the binary selection process is related to the incompleteness of the ESO-LV catalog. A significant fraction of binaries in our sample has the secondary galaxy fainter than $B_T = 15.0$. For these pairs the separation tends to be larger than pairs with brighter secondary galaxies. We associate this effect to the incompleteness of the basic catalog resulting in a surface density of faint objects much less than what would be expected from a complete catalog in the whole magnitude range. The following procedure was used to minimize such effect. For a uniform distribution of galaxies the number of objects with magnitudes between $m$ and $m + \Delta m$ can be expressed as $\Delta N \propto 10^{0.6\,m}$, and the mean separation between a primary and the nearest object is $\langle \theta \rangle = 1/2\sqrt{n}$, where $n$ is the surface density for galaxies with magnitudes in the previously defined interval. Expressing the mean separation in terms of the magnitude of the secondary galaxy we have $\langle \theta \rangle \propto 10^{-0.3\,m}$, from which follows

$$\log(\theta) = \log(\theta_\circ) - 0.3(\mathrm{m} - \mathrm{m}_\circ). \tag{4}$$

This relation predicts the relation between $\theta$ and $m$ for a random sample. When physical association is present then the separation tends to be smaller than that predicted by eq. (4). Therefore, in a diagram $\theta$ versus $m$ binary sistems tend to locate below the relation expressed by eq. (4).

Figure 2 shows the separation of the binaries in our sample, in arcsec, as a function of the magnitude of the secondary galaxy. Considering only the secondary galaxies in the interval $12.5 < \mathrm{B_T} < \mathrm{m_\circ}$, where $m_\circ \simeq 14.5$ is the limiting magnitude of the ESO-LV catalog we fitted the relation

$$\log(\theta_{\max}) = 3.254 - 0.3(\mathrm{m} - \mathrm{m}_\circ)$$

The previous fitting was done taking for each bin of 0.5 mag the three largest separations, and determining the upper boundary to the points. With such procedure we expect to have a fair representation of the separation of galaxies distributed uniformly as a function of the magnitude. The upper limit of the separation of galaxies in a uniform field represents the maximum separation binaries can reach on average, justifying why we substituted $\theta$ by $\theta_{max}$ in the last equation. We have extrapolated the fitting function to $\mathrm{m} > \mathrm{m}_\circ$ in order to eliminate those binary pairs for which $\mathrm{m_{sec}} > \mathrm{m}_\circ$ and $\log(\theta) > \log(\theta_{\max})$. Applying such restriction we were able to eliminate 106 pairs out of the 551, ending up with 445 binaries.

We have then inspected a field of $2° \times 2°$ around each binary in the sample, using the ESO-LV catalog to construct a map of the region. All pairs with probable companions inside a circle of 1 Mpc around the center of the pair were eliminated. We have used $\mathrm{H}_\circ = 50$ km/s/Mpc throughout this paper. Particular attention was paid to those cases where loose groups were easily identified in the outskirts of the 1 Mpc circle. Avoiding all these spurious cases we have defined our master list of binary sistems with 189 pairs, which are presented in Table 1. In each entry we list the NGC/IC number, or the ESO identification code, equatorial position for 1950 (RA and DEC), magnitude, numerical morphological type and velocity, for the two components of the pair. Also listed is the separation, in arcsec, and the probability parameter $p$ given by the equation 3. The last column lists the categorization of the vicinity of the pair based on the visual inpection of the maps. For those pairs with at least one available radial velocity we made a circle of 1 Mpc radius centered on the candidate pair. Two main categories were defined: (a) Probable pairs (P) – pairs with no companions inside the 1 Mpc circle, but with some suspicious companion in the neighbourhood. By suspicious we mean a galaxy with magnitude comparable to that of the pair's; (b) Isolated pairs (IP) – binary systems with no companions inside the 1Mpc circle, and also no clear companions in the neighbourhood. A probable pair with separation comparable to their component diameters was classified as CLP (Close Pair). Similarly for the isolated pairs, CLIP (Close Isolated Pair). We noticed that those pairs classified as isolated (IP+CLIP) are less frequent than the probable pairs (P+CLP).

The presence of companions in the outskirts of the 1Mpc circle was coded as PC1 (Probable Companion Outside 1 Mpc). Sometimes it is possible to see Faint Companions (FC) close to the primary, or secondary member. Finally, the presence of a nearby group outside the 1 Mpc region was coded as (NG).

It is interesting to note that even discarding pairs in dense regions the number of pairs close to or superposed on larger structures such as loose groups represent 46% of the final sample of selected binaries. This percentage is fully consistent to what it is expected from simulations (see Mamon 1993). A cross check with ZCAT has shown that among the binaries with available redshift for both galaxies 20% are false pairs ($\Delta V > 1000$km/s). This number can be regarded as an estimate of the contamination by spurious pairs in our sample.

In Figure 3 we present the results concerning the completeness of our final sample based on the $\langle V/V_{\max} \rangle$ test (Schmidt, 1968). Figure 3a shows the result for a sample of "field" galaxies extracted from the ESO-LV catalog, all of them with $b_{II} > 30°$. By "field" we mean a sample of objects with known radial velocity and located in regions where the mean local projected density of galaxies is less than 0.7 galaxies per square degree, defining a field sample that is approximately 10% of the whole ESO-LV catalog. In Figure 3a filled and open symbols represent the field sample with magnitudes corrected and not corrected for Galactic extinction, respectively. We can observe that in this case most of the "field" sample have $\langle V/V_{\max}\rangle < 0.5$. Applying the Galactic extinction correction of 0.12 cossec | b | most of the "field" sample have $\langle V/V_{\max}\rangle > 0.5$ and we can clearly see the gradual incompleteness starting at $B_T \simeq 14.5$. Application of the same methodology to the binary sample is shown in Figure 3b. It is clearly seen that in this case we have a complete sample up to $B_T \simeq 13.5$. For $m_{lim} > 13.5$ we start losing pairs in our sample because it is not possible to select primaries with companions fainter than the limiting magnitude of the catalogue. In Figure 3b the solid triangles refer to the results of the $\langle V/V\text{max}\rangle$ test taking all the binaries for which redshift is available for at least one component and we then assumed that both are at the same distance. We also show in Figure 3b, as a solid square the $\langle V/V\text{max}\rangle$ test taking only those pairs for which there is redshift available for both components and $\Delta V < 1000$km/s.

## 3 Luminosity Function of Binary Galaxies

The luminosity function (LF) was determined following a similar procedure used by Mendes de Oliveira & Hickson (1991) for compact groups. For a magnitude limited sample we have to normalize the contribution of each pair by its effective volume which is defined in a such way that the binary put at a given maximum distance would still be included in the sample. We have used the same completeness function as used by Hickson, Kindl & Auman (1989). Therefore, the expressions used to estimate the effective volume and the LF itself were those presented in Mendes de Oliveira & Hickson (1991). As discussed in the previous section, using the $\langle V/V_{\max}\rangle$ test we decided to use a limiting magnitude of 14.75. Also we restricted our samples to those pairs for which we know the redshift of at least one member and then assumed that both galaxies are at the same distance. Only pairs with galactic latitude | b |≥30° were included in the LF estimation and the galactic extinction correction was assumed to be a 0.12 cossec | b | law. The interval of absolute magnitude adopted to estimate the LF was 0.5.

Figure 4 shows the LF of the binary galaxies in our sample. For comparison we also plot the LF obtained by Xu & Sulentic (1991, hereafter XS). It is important to remind that these authors measured the LF in a different passband and using a different procedure. Apart of an offset it is quite remarkable the agreement, even confirming a small dip for $M_B$ between $-20$ and $-18$. We fit the LF with the Schechter function using a nonlinear least-square fit program (GAUSSFIT) described in Jefferys *et al.* (1988). The best fit values for the three parameters $\alpha$, $\phi^*$, and $M_*$ are $-1.47 \pm 0.16$, $4.8 \pm 2.2 \times 10^{-5}$, and $-21.23 \pm 0.20$ taking all the observed points of the LF into account; and $-1.36 \pm 0.07$, $7.0 \pm 1.7 \times 10^{-5}$, and $-21.07 \pm 0.16$ excluding the two brightest points of the LF. We also superposed on this plot the LF for field galaxies obtained by Efstathiou, Ellis, & Peterson (1988) with $\alpha = -1.07 \pm 0.05$, $\phi^* = 1.95 \pm 0.34 \times 10^{-3}$, and $M_* = -21.18 \pm 0.10$. It is interesting to note that paired galaxies represent $\sim$10% of field galaxies, corroborating previous results from XS. However, while XS found that this percentage is approximately constant over the entire luminosity range we can only confirm such trend for pairs brighter than $M_B = -19$. We should emphasize that both works suffer from serious incompleteness problems at the faint end of the LF, which in part can explain the discrepancy.

An estimate of the luminosity density of galaxies in binaries can be done using $\alpha$, $\phi^*$ and $M_*$ obtained by fitting the LF. Using the parameters obtained with all the observed points of the LF we estimate $\mathcal{L} = L^* \phi^* \Gamma(\alpha + 2) = 3.9 \times 10^6$ $L_\odot$ Mpc$^{-3}$. The luminosity density of field galaxies, as estimated by Efstathiou, Ellis, & Peterson (1988), is $0.96 \times 10^8$ $L_\odot$ Mpc$^{-3}$, implying that binary galaxies represent only 4% of the field galaxies. This low percentage compared to the value obtained by XS it is probably due to the fact that we are taking a different field LF.

## 4 Discussion

We have used the ESO-LV to build a homogeneous sample of binary systems in the southern sky. The catalogue consists of 621 double galaxies, subsequently reduced, by means of more restrictive criteria, to a sample of 189 isolated pair of galaxies which are presented in tabular form with the galaxy relevant parameters. This is the first objective catalog of binary systems in the southern sky and its general characteristics are very similar to those previously found for CPG and CMG.

The mean separation of pairs in our sample is about 6.9 times the mean diameter of both components, and 4% of the pairs in our final sample have the ratio of diameters larger than 2. This results is quite comparable to what is found in other catalogues (Sulentic 1992). Using those pairs with at least one radial velocity we estimate that their mean separation is 143 Kpc. Therefore our list contains contains more wider separation pairs than the CPG where the mean separation is 83 Kpc, and is more comparable with the CMG where the mean separation is about 146 Kpc (Soares, 1989).

The mean difference in apparent magnitude between components of CPG pairs is 0.7±0.3 while in our data this quantity is $0.97 \pm 0.67$. We conclude that algoritm has identified pairs with slightly larger difference in magnitude between the primary and secondary objects, although the difference is within 1-$\sigma$.

About 19% of objects in our sample are E or S0, while 81% are spirals. Specifically, our sample have 3% of EE pairs, 32% of ES pairs and 65% of SS pairs, close to what is expected from a sample where the objects are randomly associated, which gives 4%EE, 30%ES and 66%SS pairs. Therefore, our data does not support any evidence of overrepresentation of pairs with concordant morfological type.

We are presently conducting a redshift survey of galaxies in our sample in order to produce a homogeneous sample of pairs that allow us to study the physics of interactions based on a more representative sample than what it is presently available.

*Acknowledgments* — We thank Dr. Steve Zepf for useful suggestions. This work was supported in part by FAPESP (RRdC) Proc. No. 92/2686-9, and CNPq (DSLS) Proc. No. 300193/90-4. The list of objects in Table I is available via an anonymous FTP copy at CDS.

# APPENDIX

## The CMG Method of Finding Bound Pairs of Galaxies

In this appendix we reproduce the basic of the method devised by T.S. van Albada (see Soares 1989, chapter 2, by van Albada and Soares) to identify pairs of galaxies which exhibit physical association.

The probability of finding the first neighbour of a given galaxy, in the plane of the sky, for an ensemble of non-physically associated galaxies, is

$$P_1(\theta)d\theta = e^{-\pi\theta^2 n}\, 2\pi\theta n d\theta \quad , \tag{A1}$$

where $n$ is the surface density of galaxies. From eq. (A1), one can calculate the expected mean value of the nearest neighbour projected separation, i.e., $\langle\theta_1\rangle = 1/(2\sqrt{n})$. Defining the auxiliar variable

$$x = \frac{\theta}{\langle\theta_1\rangle} \tag{A2}$$

and inserting it in eq. (A1) we find the frequency distribution of $x$:

$$p_1(x)dx = e^{-\frac{1}{4}\pi x^2}\, \frac{\pi}{2}x dx \quad , \tag{A3}$$

which is independent of the surface density of galaxies.

The method of identifying bound pairs consists in comparing the actual distribution of $x$, $p_\circ(x)$, with $p_1(x)$. At small $x$ one expects to find an excess of $p_\circ$ over $p_1$ corresponding to physically associated pairs, while at large $x$, the distributions should coincide.

The surface density of galaxies in the neighbourhood of a given galaxy is evaluated from the expected mean separation to a neighbour of a higher order. For the tenth neighbour, $\langle\theta_{10}\rangle = \Gamma(10.5)/(9!\sqrt{\pi n})$, and one can approximate

$$\sqrt{n} \simeq \frac{\Gamma(10.5)}{9!\sqrt{\pi}\theta_{10}} = \frac{1.76}{\theta_{10}} \quad . \tag{A4}$$

The normalized projected separation to the nearest neighbour can now be calculated from $\theta_1$ and $\theta_{10}$:

$$x = \theta_1 2\sqrt{n} = 3.52\frac{\theta_1}{\theta_{10}}. \tag{A5}$$

**Figure Captions:**

**Figure 1 :** We plot the excess of probability association (p) versus the dimensionless distance (x), discriminated for different morphological types. The solid line is the fitting function for x<0.75. We can observe that the excess of probability is almost independent of the magnitude difference between the primary and secondary component. The symbols to represent such difference were used as follows: Solid circle - $\Delta J = 0$; Solid square - $\Delta J = 1$; Solid triangle - $\Delta J = 2$; Open circle - $\Delta J = 3$; Open square - $\Delta J = 4$; and Open triangle - $\Delta J = 5$.

**Figure 2 :** Separation in arcsec versus the magnitude of the secondary ($m_{sec}$). In a complete catalog we expect that all pairs should be located below the continous line. For those pairs with $m_{sec}$ fainter than 14.5 there is a tendency for higher separations, which is a natural bias produced by the incompleteness of our catalog.

**Figure 3 :** The $<V/V_{max}>$ tests for the pairs in our sample. Panel (a) refers to a sample of field galaxies extracted from the ESO-LV catalog, with filled and open symbols representing magnitudes corrected and not corrected for Galactic extinction, respectively. Panel (b) shows the same test applied to our sample of binary galaxies. Solid squares represent those pairs for which redshfit is available for both components and solid triangles when redshift is available for at least one component. The dotted line indicates the value of $<V/V_{max}>$ for a complete sample.

**Figure 4 :** The optical luminosity function of the binary galaxies in our sample (solid circle). Open circle exhibit the luminosity function obtained by Xu & Sulentic (1991). We also plot the LF for field galaxies (dashed line) as provided by Efstathiou, Ellis, & Peterson (1988). The solid line shows the best fitting Schechter function obtained using all the available points. The dotted line is the best fitting line when we remove the two brightest points of the luminosity function.

| Pair | Ident | RA | Dec | $B_T$ | T | Vel | Ident | RA | Dec | $B_T$ | T | Vel | Sep | Prob | Com |
|---|---|---|---|---|---|---|---|---|---|---|---|---|---|---|---|
| 6 | 193 0370 | 001052 | −493824 | 14.63 | −5.0 | 16400 | 193 0360 | 001046 | −493712 | 14.98 | 0.0 | 10300 | 92 | 0.95 | CLIP |
| 8 | 194 0040 | 001713 | −513242 | 14.52 | −1.0 | 6603 | 194 0041 | 001721 | −513151 | 15.17 | −0.8 | | 88 | 0.97 | CLIP |
| 9 | 539 0131 | 002225 | −210043 | 14.39 | 9.6 | | 539 0132 | 002216 | −210046 | 15.70 | 0.3 | | 124 | 0.71 | CLIP,FC |
| 10 | 350 0150 | 002302 | −331924 | 14.27 | −5.0 | 14940 | 350 0160 | 002303 | −332448 | 15.49 | 3.0 | | 324 | 0.82 | IP |
| 11 | N 0119 | 002435 | −571518 | 14.14 | −2.5 | 7340 | 150 0070 | 002314 | −572800 | 15.28 | 1.0 | | 1005 | 0.71 | P,PC1 |
| 13 | 194 0240 | 002902 | −495154 | 14.54 | 3.0 | | 194 0250 | 002933 | −495812 | 16.21 | 7.0 | | 482 | 0.60 | IP |
| 15 | 079 0030 | 002947 | −643142 | 13.78 | 3.0 | 2592 | 079 0020 | 002946 | −644000 | 14.57 | 7.1 | 2772 | 498 | 0.69 | CLIP |
| 16 | N 0148 | 003148 | −320342 | 13.11 | −1.7 | 1897 | I 1554 | 003040 | −323206 | 13.61 | −1.0 | 1806 | 1909 | 0.74 | P,PC1 |
| 18 | 242 0170 | 003254 | −442124 | 14.71 | 1.0 | 7502 | 242 0160 | 003247 | −441536 | 14.93 | −3.0 | | 356 | 0.71 | P,PC1 |
| 20 | 350 0380 | 003426 | −334954 | 14.31 | 0.7 | 6156 | 350 0402 | 003519 | −335852 | 15.26 | 7.5 | | 850 | 0.58 | P,FC |
| 21 | 242 0180 | 003444 | −465506 | 14.00 | 6.0 | | 242 0200 | 003540 | −464736 | 14.19 | 8.0 | | 729 | 0.67 | P,PC1 |
| 23 | N 0175 | 003452 | −201242 | 12.95 | 3.0 | 3930 | 540 0030 | 003310 | −202406 | 13.92 | 3.0 | 3351 | 1589 | 0.63 | P,PC1 |
| 24 | I 1562 | 003606 | −243300 | 13.56 | 5.0 | 3633 | I 1561 | 003604 | −243654 | 14.92 | 3.0 | 3886 | 235 | 0.73 | CLP,PC1 |
| 31 | 295 0100 | 004732 | −395436 | 14.18 | 1.0 | | 295 0090 | 004647 | −394842 | 15.30 | 6.0 | | 627 | 0.60 | IP |
| 34 | N 0319 | 005439 | −440630 | 14.25 | 0.0 | | N 0322 | 005452 | −435948 | 14.29 | −2.0 | | 425 | 0.86 | P,PC1 |
| 40 | N 0348 | 005841 | −533048 | 14.58 | 3.0 | | 151 0180 | 005925 | −532806 | 14.80 | 1.0 | | 424 | 0.65 | P,N |
| 42 | I 1615 | 010156 | −512400 | 14.29 | 4.0 | 7670 | I 1617 | 010206 | −511800 | 14.59 | −1.0 | 7970 | 371 | 0.70 | P,PC1 |
| 43 | 295 0380 | 010454 | −421100 | 14.08 | −3.0 | | 295 0370 | 010450 | −421630 | 15.53 | 3.0 | | 332 | 0.89 | P,PC1 |
| 45 | N 0418 | 010814 | −302912 | 13.19 | 5.0 | 5684 | I 1637 | 010839 | −304212 | 13.59 | 5.0 | 6002 | 844 | 0.63 | P,PC1 |
| 57 | 151 0361 | 011217 | −553953 | 13.11 | 4.5 | | N 0454 | 011220 | −553942 | 13.13 | 1.0 | 3627 | 27 | 0.80 | CLP,PC1 |
| 59 | 244 0121 | 011556 | −444319 | 14.43 | −0.3 | | 244 0120 | 011556 | −444336 | 15.70 | 3.0 | 6700 | 17 | 0.99 | CLP,N |
| 65 | 542 0150 | 012450 | −220154 | 14.53 | −2.0 | 5567 | 542 0160 | 012514 | −215406 | 15.83 | 2.6 | | 574 | 0.88 | IP |
| 66 | 113 0500 | 012740 | −614318 | 13.90 | −5.0 | | 113 0490 | 012626 | −614148 | 15.47 | 0.5 | | 533 | 0.78 | IP |
| 67 | 476 0160 | 012805 | −270218 | 14.29 | 4.5 | 6021 | 476 0180 | 012854 | −270700 | 14.49 | −2.0 | | 712 | 0.65 | P,PC1 |
| 75 | N 0633 | 013411 | −373442 | 13.50 | 2.8 | 5160 | 297 0120 | 013411 | −373548 | 15.06 | −5.0 | | 66 | 0.77 | CLIP,N |
| 78 | 297 0180 | 013627 | −401554 | 14.28 | 1.0 | | 297 0160 | 013549 | −401924 | 14.73 | 6.0 | | 482 | 0.65 | P,PC1 |
| 79 | N 0642 | 013649 | −301006 | 13.58 | 5.0 | 5883 | N 0639 | 013641 | −301042 | 14.67 | 1.0 | 5826 | 109 | 0.77 | CLIP |
| 83 | 003 0030 | 014021 | −833700 | 14.74 | 6.0 | | 003 0040 | 014306 | −832748 | 15.02 | 4.0 | | 618 | 0.65 | IP |
| 85 | 353 0400 | 014132 | −362024 | 13.61 | −1.0 | 5304 | 353 0410 | 014251 | −362206 | 14.93 | 0.0 | 5378 | 959 | 0.59 | IP |
| 86 | 297 0230 | 014220 | −405454 | 14.46 | 3.0 | 10121 | 297 0240 | 014229 | −404912 | 15.91 | 3.0 | | 356 | 0.65 | P,PC1 |
| 87 | 114 0070 | 014446 | −585518 | 14.22 | 9.0 | | 114 0071 | 014445 | −585520 | 14.44 | 7.3 | | 8 | 0.75 | CLP,FC |
| 88 | 013 0210 | 014522 | −781148 | 14.65 | 1.0 | | 013 0200 | 014339 | −781236 | 16.14 | 10.0 | | 319 | 0.70 | IP |
| 90 | N 0696 | 014718 | −350912 | 14.37 | −1.0 | 8075 | N 0698 | 014731 | −350442 | 14.78 | 2.7 | 8342 | 313 | 0.82 | CLIP |
| 94 | 477 0140 | 015242 | −261548 | 14.52 | −2.0 | | 477 0150 | 015306 | −261030 | 15.80 | 3.0 | | 453 | 0.88 | P,PC1 |
| 95 | I 1759 | 015543 | −331348 | 13.82 | 5.0 | 3853 | I 1762 | 015536 | −332900 | 14.36 | 5.0 | 5683 | 916 | 0.50 | P,PC1 |
| 96 | I 1763 | 015655 | −280306 | 14.70 | 3.0 | | 414 0180 | 015626 | −281130 | 15.75 | −1.4 | | 633 | 0.63 | IP |
| 99 | 114 0210 | 020021 | −584842 | 14.19 | 5.0 | | 114 0190 | 020006 | −583706 | 14.68 | 1.0 | | 705 | 0.63 | P,PC1 |
| 100 | N 0822 | 020436 | −412342 | 14.25 | −5.0 | 5395 | 298 0080 | 020413 | −414536 | 14.72 | 7.0 | 5397 | 1339 | 0.65 | P,PC1 |
| 101 | 052 0200 | 020453 | −712112 | 14.58 | 4.0 | | 052 0210 | 020507 | −712200 | 14.97 | 3.2 | | 82 | 0.78 | CLIP |
| 102 | 478 0060 | 020700 | −233906 | 13.22 | 4.0 | | 478 0070 | 020712 | −233306 | 16.07 | 8.0 | | 395 | 0.69 | IP |
| 104 | 298 0160 | 020850 | −393600 | 13.83 | 1.0 | 5205 | 298 0190 | 020952 | −392624 | 14.31 | 5.0 | 5391 | 920 | 0.53 | P,PC1 |
| 105 | N 0858 | 021011 | −224218 | 14.19 | 5.0 | 12356 | 478 0140 | 021046 | −224330 | 15.86 | 1.0 | | 489 | 0.64 | P,N |
| 106 | 153 0290 | 021231 | −545506 | 14.67 | 1.0 | | 153 0300 | 021246 | −545800 | 14.95 | 4.6 | | 216 | 0.71 | CLP,PC1 |
| 109 | N 0888 | 021559 | −600536 | 14.47 | −5.0 | | 115 0030 | 021628 | −595700 | 15.96 | 1.0 | | 559 | 0.78 | P,PC1 |
| 112 | 115 0080 | 022253 | −583718 | 14.65 | −5.0 | 9246 | 115 0090 | 022300 | −583936 | 15.15 | 2.0 | | 148 | 0.95 | CLP,N |

| Pair | Ident | RA | Dec | $B_T$ | T | Vel | Ident | RA | Dec | $B_T$ | T | Vel | Sep | Prob | Com |
|---|---|---|---|---|---|---|---|---|---|---|---|---|---|---|---|
| 116 | 198 0130 | 022728 | −484242 | 13.66 | 2.0 | 6200 | 198 0140 | 022748 | −485112 | 14.98 | 1.0 | 10500 | 546 | 0.65 | IP |
| 117 | I 1813 | 022843 | −342630 | 14.20 | −1.0 | 4483 | I 1811 | 022832 | −342906 | 14.35 | 2.0 | 4821 | 206 | 0.95 | CLIP |
| 125 | 154 0100 | 024340 | −555700 | 13.39 | 0.0 | 5507 | 154 0130 | 024448 | −554000 | 13.61 | 3.0 | 6248 | 1170 | 0.90 | IP |
| 126 | 416 0180 | 024355 | −322148 | 14.39 | −2.0 | | 416 0170 | 024349 | −321530 | 15.60 | 3.0 | | 385 | 0.89 | CLP,N |
| 130 | 479 0400 | 024429 | −253324 | 14.64 | 3.5 | 10548 | 479 0410 | 024436 | −253554 | 16.01 | 0.5 | | 177 | 0.75 | CLP,N |
| 132 | 356 0130 | 024647 | −365518 | 14.52 | 6.0 | | 356 0120 | 024643 | −364324 | 15.82 | 5.4 | | 715 | 0.60 | IP |
| 135 | N 1136 | 024925 | −551048 | 13.80 | 2.0 | 5438 | N 1135 | 024918 | −550800 | 16.16 | 7.0 | 13339 | 178 | 0.73 | CLIP |
| 139 | 356 0220 | 025550 | −365500 | 13.30 | 4.0 | 6170 | 356 0240 | 025656 | −364836 | 14.83 | 3.4 | 6124 | 880 | 0.61 | P,PC1 |
| 147 | 481 0090 | 031007 | −244830 | 14.68 | 1.0 | | 481 0110 | 031014 | −244400 | 15.44 | 3.0 | | 286 | 0.64 | CLP,N |
| 148 | 417 0210 | 031113 | −315018 | 14.13 | −3.0 | 4125 | 417 0200 | 031045 | −314024 | 15.14 | 6.0 | | 693 | 0.84 | P,PC1 |
| 151 | I 1908 | 031343 | −550006 | 14.53 | 3.0 | 8234 | 155 0131 | 031344 | −550017 | 16.14 | 3.0 | | 12 | 0.79 | CLIP,N |
| 167 | 418 0071 | 032753 | −285708 | 14.09 | 4.5 | | 418 0070 | 032752 | −285642 | 14.94 | −2.0 | 11052 | 29 | 0.79 | CLIP |
| 200 | 549 0360 | 035253 | −173648 | 14.46 | 4.0 | 8501 | 549 0300 | 035138 | −174424 | 14.92 | 3.0 | | 1164 | 0.50 | IP |
| 207 | N 1512 | 040216 | −432912 | 11.08 | 1.0 | 896 | N 1510 | 040154 | −433212 | 13.47 | −2.0 | 1004 | 299 | 0.75 | CLIP |
| 215 | N 1534 | 040807 | −625536 | 13.76 | 0.1 | | N 1529 | 040641 | −630154 | 14.39 | −2.0 | | 697 | 0.68 | P,PC1 |
| 221 | 420 0141 | 041310 | −283611 | 14.46 | −1.6 | | N 1540 | 041309 | −283624 | 14.75 | 10.0 | | 15 | 0.99 | P,C |
| 222 | N 1558 | 041843 | −450900 | 13.28 | 4.0 | 4541 | 250 0180 | 041927 | −451124 | 15.11 | 3.0 | | 487 | 0.66 | CLP,PC1 |
| 223 | N 1567 | 041943 | −482218 | 13.36 | −5.0 | 4552 | 202 0090 | 041938 | −482524 | 15.32 | 5.0 | 4681 | 192 | 0.94 | CLIP |
| 227 | I 2073 | 042522 | −531754 | 14.39 | 6.0 | 3989 | 157 0282 | 042458 | −531552 | 16.04 | 5.6 | | 247 | 0.68 | CLP,FC,N |
| 238 | 158 0070 | 044831 | −535948 | 14.64 | 3.0 | 15100 | 158 0060 | 044734 | −535954 | 15.34 | 0.0 | | 502 | 0.60 | IP |
| 240 | 203 0070 | 044930 | −475106 | 14.31 | −3.0 | | N 1680 | 044710 | −475412 | 14.45 | 3.8 | | 1420 | 0.69 | IP |
| 244 | 085 0141 | 045408 | −625255 | 12.76 | 8.3 | | 085 0140 | 045414 | −625242 | 13.42 | 9.4 | 1119 | 41 | 0.74 | CLP,N |
| 247 | 552 0400 | 045639 | −213836 | 14.30 | 2.0 | 6818 | 552 0320 | 045601 | −214136 | 14.39 | 3.0 | | 559 | 0.59 | P,PC1 |
| 250 | 361 0250 | 045952 | −340612 | 14.26 | 3.1 | 5287 | 362 0010 | 050006 | −340606 | 15.08 | 0.0 | | 173 | 0.77 | CLP,PC1 |
| 255 | 486 0230 | 050156 | −240354 | 14.16 | −3.0 | 12440 | 486 0180 | 050105 | −240418 | 15.57 | 1.0 | | 698 | 0.62 | P,PC1 |
| 256 | 085 0340 | 050256 | −634912 | 13.76 | −2.0 | | 085 0380 | 050356 | −633854 | 13.87 | 4.0 | 4845 | 735 | 0.75 | P,N |
| 259 | N 1803 | 050409 | −493800 | 13.51 | 4.0 | 4085 | 203 0190 | 050417 | −493948 | 14.07 | 1.0 | 4351 | 133 | 0.78 | CLP,PC1 |
| 261 | 033 0110 | 050615 | −734306 | 14.09 | 1.5 | | 033 0140 | 050826 | −733706 | 15.01 | 5.5 | | 659 | 0.67 | P,PC1 |
| 266 | N 1853 | 051125 | −572724 | 13.58 | 6.7 | | 158 0200 | 050952 | −572436 | 15.43 | 3.0 | | 769 | 0.66 | P,FC |
| 267 | 486 0380 | 051307 | −223618 | 14.59 | −3.0 | | 486 0400 | 051333 | −224548 | 14.76 | 3.0 | | 674 | 0.71 | IP |
| 268 | 486 0391 | 051320 | −263120 | 14.41 | 3.0 | | 486 0390 | 051320 | −263130 | 15.26 | 1.0 | 3840 | 9 | 0.80 | CLP,PC1,FC |
| 270 | 119 0470 | 051402 | −621700 | 14.07 | 3.0 | 5100 | 119 0460 | 051400 | −621330 | 15.04 | 3.0 | | 210 | 0.72 | CLP,N |
| 272 | 486 0490 | 051522 | −234754 | 14.67 | 3.0 | | 486 0440 | 051434 | −235030 | 15.29 | 1.0 | | 676 | 0.50 | P,PC1 |
| 273 | 362 0180 | 051744 | −324230 | 13.81 | −0.3 | 3790 | 362 0170 | 051736 | −324830 | 15.38 | 3.0 | | 373 | 0.90 | CLP,PC1 |
| 276 | 305 0220 | 052120 | −385506 | 14.56 | 3.0 | | 305 0210 | 052103 | −390642 | 14.62 | 3.7 | | 723 | 0.66 | P,PC1 |
| 277 | 305 0250 | 052409 | −395700 | 14.27 | 4.0 | | 305 0230 | 052311 | −400412 | 15.44 | 1.4 | | 794 | 0.58 | P,PC1 |
| 278 | N 1930 | 052433 | −464618 | 13.39 | −5.0 | 4224 | 253 0010 | 052257 | −464724 | 14.65 | 7.0 | | 988 | 0.81 | P,N |
| 280 | 306 0030 | 052729 | −392736 | 14.44 | 4.0 | | 306 0011 | 052625 | −393641 | 15.37 | 0.7 | | 918 | 0.53 | P,N |
| 281 | N 2008 | 053352 | −505954 | 14.64 | 5.0 | 10341 | N 2007 | 053347 | −505712 | 14.84 | 6.0 | 4523 | 168 | 0.75 | CLP,PC1,N |
| 284 | 568 0090 | 102405 | −194712 | 14.28 | 6.0 | 3108 | 568 0080 | 102344 | −195848 | 14.65 | −5.0 | | 756 | 0.56 | IP |
| 288 | 569 0240 | 105433 | −195406 | 12.89 | 4.0 | | 569 0270 | 105524 | −194400 | 14.09 | 1.0 | | 940 | 0.61 | P,FC |
| 289 | N 3511 | 110057 | −224900 | 11.66 | 5.3 | 1104 | N 3513 | 110119 | −225830 | 12.16 | 5.3 | 1195 | 645 | 0.74 | CLP |
| 304 | 505 0150 | 120433 | −252448 | 14.04 | −4.0 | 7541 | 505 0170 | 120455 | −253536 | 15.77 | 0.0 | | 713 | 0.82 | IP |
| 307 | I 3152 | 121700 | −255206 | 13.28 | −3.5 | 3275 | 506 0020 | 121734 | −254724 | 14.60 | 4.3 | 3960 | 538 | 0.90 | CLP,PC1 |
| 311 | 574 0240 | 124055 | −203406 | 14.69 | 2.0 | 6810 | 574 0230 | 124055 | −202812 | 15.51 | 2.0 | | 354 | 0.64 | CLP,PC1,N |

| Pair | Ident | RA | Dec | $B_T$ | T | Vel | Ident | RA | Dec | $B_T$ | T | Vel | Sep | Prob | Com |
|---|---|---|---|---|---|---|---|---|---|---|---|---|---|---|---|
| 316 | 507 0450 | 125254 | −263312 | 12.84 | −2.0 | 4875 | 507 0460 | 125303 | −263218 | 13.96 | −3.5 | 4602 | 132 | 0.95 | CLP,N |
| 318 | 443 0220 | 125740 | −311006 | 14.48 | −2.0 | 3710 | 443 0270 | 125822 | −311000 | 15.97 | 7.7 | | 539 | 0.83 | CLP,N |
| 319 | 575 0441 | 125741 | −222534 | 13.53 | −5.0 | | 575 0440 | 125739 | −222524 | 15.06 | −6.0 | | 31 | 0.98 | CLP,N |
| 323 | 443 0500 | 130139 | −282600 | 14.49 | −2.0 | | 443 0510 | 130140 | −282842 | 15.61 | 3.0 | | 162 | 0.92 | CLP,N |
| 331 | 444 0101 | 131743 | −303802 | 14.33 | 0.8 | | 444 0100 | 131741 | −303848 | 15.43 | 6.0 | 9211 | 51 | 0.88 | CLP,N |
| 337 | N 5134 | 132236 | −205224 | 12.11 | 1.0 | 1757 | I 4237 | 132150 | −205236 | 13.25 | 4.0 | 2643 | 644 | 0.67 | P,PC1 |
| 340 | 576 0760 | 132759 | −220948 | 13.86 | −3.6 | | 576 0730 | 132753 | −221506 | 15.11 | 6.0 | | 328 | 0.79 | P,N |
| 344 | 444 0760 | 133301 | −300736 | 14.35 | −1.0 | | 444 0770 | 133303 | −301036 | 15.26 | 8.0 | 3847 | 181 | 0.89 | CLP,N |
| 348 | N 5260 | 133734 | −233618 | 13.63 | 5.0 | 6539 | 509 0930 | 133740 | −234006 | 15.94 | 2.0 | | 242 | 0.64 | CLP,N |
| 350 | I 4320 | 134115 | −265848 | 14.25 | −1.0 | 6827 | 509 1000 | 134103 | −270636 | 15.13 | 3.0 | 6567 | 494 | 0.75 | IP |
| 351 | 445 0510 | 134631 | −275706 | 14.60 | −1.0 | 4995 | 445 0480 | 134618 | −274524 | 15.58 | 7.4 | | 722 | 0.69 | P,PC1 |
| 355 | I 4350 | 135425 | −250006 | 13.70 | 2.0 | 6157 | 510 023 | 135427 | −250842 | 15.15 | −2.0 | | 516 | 0.50 | P,N |
| 358 | 510 0590 | 140157 | −243518 | 13.49 | 6.0 | 2337 | 510 0580 | 140148 | −243536 | 14.06 | 6.0 | 2333 | 124 | 0.77 | CLP,N |
| 361 | 578 0260 | 140554 | −212142 | 14.33 | 4.0 | 2781 | 578 0300 | 140645 | −212230 | 15.73 | 5.5 | | 714 | 0.57 | P,PC1 |
| 362 | 578 0290 | 140643 | −173748 | 14.31 | −5.0 | | 578 0320 | 140730 | −173606 | 15.08 | 2.0 | | 679 | 0.87 | IP |
| 363 | 511 0180 | 141449 | −235712 | 14.51 | 3.0 | | 511 0200 | 141459 | −235724 | 15.57 | 3.0 | | 137 | 0.75 | CLP,PC1 |
| 373 | 580 0430 | 144810 | −181600 | 13.79 | −2.9 | 6054 | 580 0410 | 144748 | −175642 | 14.32 | 3.0 | | 1199 | 0.82 | IP |
| 375 | I 4536 | 151023 | −175706 | 13.71 | 7.0 | 2279 | N 5863 | 150758 | −181430 | 13.72 | 1.0 | 4266 | 2316 | 0.58 | P,FC |
| 378 | I 4901 | 195012 | −585042 | 12.29 | 5.0 | 2122 | 142 0510 | 195047 | −585148 | 14.72 | 2.0 | 11264 | 279 | 0.70 | CLP,N |
| 387 | I 4935 | 200029 | −574424 | 14.04 | 0.8 | | 143 0040 | 200041 | −574918 | 15.19 | 5.0 | | 309 | 0.74 | P,N |
| 388 | I 4938 | 200157 | −602112 | 13.33 | 2.3 | 3587 | 143 0050 | 200053 | −601824 | 15.77 | −3.0 | | 504 | 0.60 | CLP,PC1 |
| 392 | 233 0370 | 200559 | −492842 | 14.56 | 3.0 | | 233 0360 | 200554 | −492212 | 14.77 | 2.0 | 3200 | 393 | 0.62 | CLP,N |
| 399 | 400 0050 | 201336 | −370830 | 14.08 | 5.0 | 6085 | 400 0020 | 201303 | −371012 | 15.75 | 5.6 | | 407 | 0.68 | CLP,PC1 |
| 402 | 340 0150 | 201604 | −412918 | 13.59 | 1.0 | 6473 | 340 0140 | 201552 | −413000 | 14.74 | 1.0 | | 141 | 0.75 | CLP,N |
| 404 | 340 0170 | 201622 | −392642 | 13.19 | 8.0 | 2594 | 340 0120 | 201514 | −392936 | 13.72 | 5.0 | 2719 | 806 | 0.56 | CLP,PC1 |
| 405 | 234 0110 | 201828 | −480848 | 14.59 | −2.0 | 5702 | 234 0090 | 201814 | −480954 | 15.03 | 5.4 | | 154 | 0.92 | CLP,N |
| 407 | 462 0150 | 202011 | −275230 | 12.92 | −5.0 | 5827 | 462 0160 | 202036 | −282618 | 13.61 | 6.0 | | 2054 | 0.78 | IP |
| 409 | 234 0210 | 202042 | −495048 | 13.92 | −3.0 | 5395 | 234 0140 | 201917 | −494000 | 14.72 | −3.0 | 11700 | 1048 | 0.53 | P,N |
| 416 | I 5020 | 202729 | −333912 | 12.95 | 4.0 | 3091 | 400 0370 | 202804 | −333848 | 14.58 | 6.0 | 3202 | 437 | 0.70 | CLP,N |
| 422 | 463 0080 | 203431 | −274506 | 14.43 | 1.0 | | 463 0091 | 203435 | −274436 | 15.14 | 5.4 | | 65 | 0.78 | CLP,FC |
| 423 | N 6935 | 203439 | −521706 | 12.77 | 1.0 | 4631 | N 6937 | 203505 | −521912 | 13.37 | 4.7 | 4680 | 269 | 0.75 | CLP,FC,PC1 |
| 424 | N 6920 | 203630 | −801048 | 13.08 | −3.0 | 2774 | 026 0050 | 203835 | −801542 | 15.97 | 9.0 | | 433 | 0.90 | CLP,PC1 |
| 425 | I 5034 | 203949 | −571242 | 14.67 | 4.0 | | I 5035 | 204022 | −571830 | 15.89 | 4.7 | | 439 | 0.65 | P,PC1 |
| 426 | 463 0210 | 204031 | −295306 | 13.11 | 7.0 | 2702 | 463 0200 | 204011 | −300200 | 13.45 | 3.0 | 2701 | 593 | 0.68 | CLP,FC,PC1 |
| 427 | 074 0081 | 204042 | −674252 | 13.62 | 9.9 | | 074 0080 | 204044 | −674336 | 14.68 | 3.0 | 10213 | 44 | 0.74 | CLIP |
| 429 | I 5042 | 204324 | −651612 | 14.06 | 6.0 | 4214 | I 5038 | 204229 | −651200 | 14.22 | 6.0 | 4157 | 427 | 0.73 | CLP,PC1 |
| 432 | I 5063 | 204812 | −571530 | 12.92 | −0.4 | 3402 | I 5064 | 204848 | −572512 | 14.31 | 1.0 | 3377 | 650 | 0.86 | P,PC1 |
| 435 | 286 0160 | 205413 | −464754 | 14.62 | 2.0 | | 286 0150 | 205412 | −464542 | 16.23 | 1.4 | | 132 | 0.73 | CLIP |
| 437 | 286 0180 | 205431 | −433400 | 14.30 | 4.0 | 9162 | 286 0170 | 205430 | −433242 | 14.49 | −2.0 | | 78 | 0.77 | CLP,N |
| 439 | 598 0090 | 205537 | −201030 | 14.30 | 3.0 | | 598 0110 | 205615 | −201412 | 15.69 | 6.0 | | 579 | 0.65 | IP |
| 449 | 286 0500 | 210325 | −424524 | 13.80 | −3.0 | 2733 | 286 0440 | 210222 | −425848 | 14.37 | −1.0 | | 1061 | 0.58 | P,N |
| 455 | N 7029 | 210826 | −492918 | 12.54 | −3.0 | 2863 | N 7022 | 210608 | −493024 | 14.04 | −1.0 | 2400 | 1346 | 0.61 | P,FC,N |
| 456 | 598 0310 | 210909 | −200000 | 14.30 | −3.0 | | 598 0300 | 210858 | −195724 | 15.39 | 2.6 | | 219 | 0.92 | IP |
| 457 | 342 0220 | 210947 | −380436 | 14.60 | 3.0 | | 342 0230 | 210954 | −380536 | 15.31 | 6.0 | | 102 | 0.76 | CLP,N |
| 460 | I 5101 | 211745 | −660300 | 14.04 | 4.0 | 5166 | I 5100 | 211733 | −660848 | 14.61 | 5.0 | | 355 | 0.71 | CLP,PC1 |

| Pair | Ident | RA | Dec | $B_T$ | T | Vel | Ident | RA | Dec | $B_T$ | T | Vel | Sep | Prob | Com |
|---|---|---|---|---|---|---|---|---|---|---|---|---|---|---|---|
| 467 | 530 0480 | 212419 | −231418 | 14.73 | −1.6 | | 530 0470 | 212411 | −231236 | 15.14 | 1.0 | 9762 | 150 | 0.95 | CLP,N |
| 473 | N 7075 | 212826 | −385018 | 13.80 | −4.0 | 5487 | 343 0030 | 212809 | −385030 | 14.25 | −2.0 | | 198 | 0.96 | CLIP |
| 474 | I 5109 | 212846 | −742006 | 14.66 | −3.9 | | I 5103 | 212413 | −741718 | 15.04 | 5.0 | | 1119 | 0.62 | P,PC1 |
| 482 | N 7110 | 213912 | −342324 | 14.25 | 4.0 | 5342 | N 7109 | 213858 | −344024 | 14.42 | −2.0 | | 1034 | 0.57 | P,PC1 |
| 485 | 288 0020 | 214303 | −464448 | 13.54 | 4.1 | 9875 | 288 0010 | 214302 | −464506 | 13.74 | 0.2 | 9588 | 20 | 0.80 | CLP,N |
| 486 | N 7124 | 214447 | −504748 | 13.28 | 4.5 | 5190 | 236 0480 | 214433 | −505330 | 15.99 | −0.5 | | 366 | 0.61 | CLP,FC,PC1 |
| 487 | N 7125 | 214537 | −605642 | 12.70 | 5.3 | 3148 | N 7126 | 214539 | −605030 | 12.99 | 4.0 | 2981 | 372 | 0.76 | CLP,FC,PC1 |
| 488 | 027 0010 | 214548 | −814554 | 12.22 | 5.0 | 2500 | 027 0030 | 214701 | −815318 | 14.81 | 10.0 | | 470 | 0.72 | CLIP,FC |
| 489 | I 5139 | 214730 | −311342 | 13.34 | −3.0 | 5362 | 466 0090 | 214709 | −310930 | 15.82 | 3.0 | | 368 | 0.90 | CLP,PC1 |
| 490 | 048 0110 | 214801 | −733154 | 14.72 | 4.0 | | 048 0120 | 214814 | −733148 | 16.25 | 5.4 | | 55 | 0.78 | CLP,N |
| 493 | I 5141 | 214943 | −594348 | 13.49 | 4.5 | 4392 | 145 2210 | 214925 | −593742 | 15.69 | 4.5 | | 390 | 0.70 | CLP,N |
| 494 | 288 0210 | 215241 | −432730 | 14.20 | 5.0 | 7849 | 288 0200 | 215221 | −432918 | 15.16 | 0.0 | | 243 | 0.73 | CLP,N |
| 496 | 404 0120 | 215409 | −344918 | 12.89 | 4.0 | 2603 | N 7154 | 215223 | −350306 | 12.98 | 7.0 | 2636 | 1544 | 0.62 | P,FC,PC1 |
| 497 | I 5149 | 215519 | −273754 | 13.84 | 3.0 | | 466 0270 | 215607 | −273912 | 14.50 | 3.0 | 5483 | 642 | 0.51 | IP |
| 500 | 288 0300 | 215828 | −424142 | 14.71 | 3.0 | | 288 0320 | 215833 | −424048 | 14.95 | 3.0 | | 77 | 0.78 | CLP,N |
| 507 | I 5154 | 220042 | −662132 | 14.29 | 3.0 | | 108 0101 | 220042 | −662130 | 15.84 | 10.0 | | 3 | 0.80 | CLIP |
| 509 | 466 0500 | 220053 | −280230 | 14.28 | −2.0 | 7104 | 466 0490 | 220050 | −281042 | 15.54 | 3.0 | | 493 | 0.64 | P,N |
| 511 | 146 2060 | 220159 | −604500 | 14.61 | −3.0 | | 146 0060 | 220146 | −605512 | 15.51 | 5.0 | | 619 | 0.83 | IP |
| 514 | N 7205 | 220510 | −574118 | 11.77 | 5.0 | 1482 | 146 0070 | 220408 | −574230 | 14.56 | 5.5 | | 502 | 0.70 | CLP,FC |
| 520 | N 7216 | 220844 | −685430 | 13.41 | −3.7 | 3459 | 076 0070 | 221015 | −684642 | 15.80 | 10.0 | | 679 | 0.72 | CLP,PC1 |
| 521 | 404 0390 | 220847 | −340800 | 14.27 | −1.0 | | 404 0391 | 220852 | −340750 | 16.23 | −3.0 | | 61 | 0.98 | CLP,N |
| 522 | 237 0450 | 220911 | −485306 | 14.20 | 1.0 | | 237 0470 | 221004 | −490536 | 14.60 | 6.0 | | 913 | 0.61 | P,N |
| 531 | 533 0070 | 221346 | −270412 | 14.34 | 2.0 | | 533 0060 | 221316 | −265548 | 14.38 | 3.0 | 7105 | 644 | 0.53 | P,N |
| 533 | N 7247 | 221454 | −235900 | 13.45 | 3.0 | | 533 0090 | 221502 | −240836 | 14.66 | −3.0 | | 586 | 0.54 | P,PC1 |
| 534 | I 5188 | 221502 | −595330 | 13.83 | 5.0 | | I 5187 | 221454 | −595130 | 14.69 | 3.0 | | 134 | 0.77 | CLP,PC1 |
| 536 | I 5203 | 221913 | −600136 | 14.74 | 5.0 | | I 5205 | 221926 | −600224 | 15.98 | 7.0 | | 108 | 0.76 | CLP,N |
| 537 | I 5202 | 221918 | −660318 | 14.67 | 5.0 | | I 5200 | 221838 | −660106 | 15.96 | 6.0 | | 277 | 0.72 | CLP,N |
| 538 | 533 0210 | 221947 | −234606 | 14.32 | −5.0 | | 533 0200 | 221946 | −234636 | 14.74 | −2.0 | | 32 | 0.99 | CLIP |
| 539 | I 5210 | 221948 | −190724 | 14.01 | −3.0 | | I 5211 | 222000 | −190800 | 14.66 | 1.0 | 7382 | 173 | 0.95 | CLIP |
| 541 | 405 0150 | 221957 | −371212 | 14.45 | −0.6 | 9950 | 405 0110 | 221837 | −371700 | 14.81 | 0.0 | | 997 | 0.58 | IP |
| 543 | I 5212 | 222035 | −381730 | 14.63 | 4.0 | 8320 | I 5209 | 222013 | −381454 | 15.58 | 2.6 | | 302 | 0.72 | CLP,PC1 |
| 546 | N 7285 | 222552 | −250548 | 12.82 | 1.0 | 4347 | N 7284 | 222550 | −250600 | 12.96 | −2.0 | 4706 | 29 | 0.80 | CLP,N |
| 549 | 533 0350 | 222721 | −270148 | 14.68 | −1.5 | | 533 0370 | 222752 | −271118 | 15.34 | 7.0 | | 704 | 0.67 | P,N |
| 555 | 027 0140 | 223059 | −811924 | 13.84 | 3.0 | | 027 0090 | 222308 | −811518 | 14.47 | 7.0 | | 1097 | 0.62 | P,PC1 |
| 556 | 405 0290 | 223113 | −323918 | 14.61 | 5.0 | | 405 0300 | 223147 | −323824 | 15.73 | −0.6 | | 432 | 0.57 | P,N |
| 558 | N 7310 | 223153 | −224436 | 14.68 | 4.3 | 9686 | 533 0480 | 223148 | −225706 | 14.69 | 3.0 | | 753 | 0.52 | IP |
| 560 | 534 0020 | 223343 | −243606 | 13.92 | −4.0 | | 534 0060 | 223436 | −243024 | 14.46 | −2.5 | | 799 | 0.71 | P,PC1 |
| 562 | 534 0130 | 223631 | −264618 | 14.32 | −3.0 | 8193 | 534 0140 | 223641 | −263948 | 15.56 | 6.0 | | 412 | 0.81 | P,N |
| 563 | 406 0040 | 223941 | −372648 | 14.30 | −1.9 | | 345 0420 | 223940 | −373430 | 15.10 | −2.0 | | 462 | 0.81 | P,N |
| 568 | N 7377 | 224505 | −223436 | 11.39 | −1.0 | 3351 | 534 0240 | 224232 | −225936 | 13.66 | 6.5 | 3113 | 2593 | 0.59 | P,PC1 |
| 573 | 406 0181 | 225127 | −372058 | 14.65 | −3.0 | | 406 0180 | 225127 | −372054 | 15.17 | −0.4 | | 4 | 0.99 | CLP,N |
| 574 | N 7410 | 225211 | −395542 | 11.35 | 1.0 | 1751 | N 7404 | 225129 | −393454 | 13.64 | −3.0 | 1899 | 1338 | 0.61 | P,FC,PC1 |
| 575 | 076 0310 | 225229 | −705024 | 13.96 | −4.0 | | 076 0300 | 225157 | −705354 | 14.20 | −0.3 | 3750 | 262 | 0.95 | CLIP |
| 580 | I 5263 | 225452 | −691918 | 14.23 | −2.0 | 3798 | I 5279 | 225944 | −692848 | 14.31 | 1.0 | 3901 | 1642 | 0.62 | IP |
| 589 | 469 0140 | 230444 | −280512 | 14.41 | 0.3 | 3779 | 469 0130 | 230432 | −275718 | 15.39 | 9.3 | | 499 | 0.79 | IP |

TABLE 1. *ESO-LV Binary list (continued)*

| Pair | Ident | RA | Dec | $B_T$ | T | Vel | Ident | RA | Dec | $B_T$ | T | Vel | Sep | Prob | Com |
|---|---|---|---|---|---|---|---|---|---|---|---|---|---|---|---|
| 594 | N 7645 | 232107 | −293942 | 13.80 | 5.0 | 6869 | N 7636 | 231953 | −293318 | 14.70 | −2.0 | | 1038 | 0.57 | P,PC1 |
| 595 | N 7637 | 232242 | −821112 | 13.24 | 4.7 | 3596 | 027 0240 | 230922 | −815354 | 14.17 | 3.3 | | 1958 | 0.51 | IP |
| 597 | 347 0171 | 232413 | −373719 | 14.19 | 8.5 | | 347 0170 | 232415 | −373718 | 15.77 | 9.4 | 690 | 21 | 0.74 | CLIP |
| 609 | 077 0300 | 234140 | −700806 | 14.59 | 6.0 | | 077 0020 | 234139 | −700806 | 14.81 | 6.3 | | 5 | 0.80 | CLIP |
| 615 | 110 0290 | 234932 | −625100 | 14.68 | 1.0 | | 078 0090 | 234949 | −625030 | 15.58 | 1.0 | | 120 | 0.76 | CLIP |
| 616 | 471 0340 | 234956 | −302730 | 14.33 | 1.0 | | 471 0320 | 234947 | −303030 | 14.74 | −2.0 | | 214 | 0.74 | CLP,FC,PC1 |
| 617 | 111 0100 | 235331 | −605700 | 14.18 | 4.0 | 4829 | 111 0090 | 235319 | −605736 | 14.65 | 5.0 | 4279 | 94 | 0.78 | CLP,PC1 |
| 618 | 349 0100 | 235426 | −350212 | 13.89 | −4.0 | 14696 | 349 0090 | 235426 | −345730 | 14.64 | 3.0 | 12489 | 282 | 0.93 | CLP,PC1 |
| 621 | 349 0170 | 235824 | −335324 | 14.46 | 6.0 | 6979 | 349 0200 | 235930 | −334448 | 14.71 | 1.0 | 8755 | 970 | 0.51 | IP |

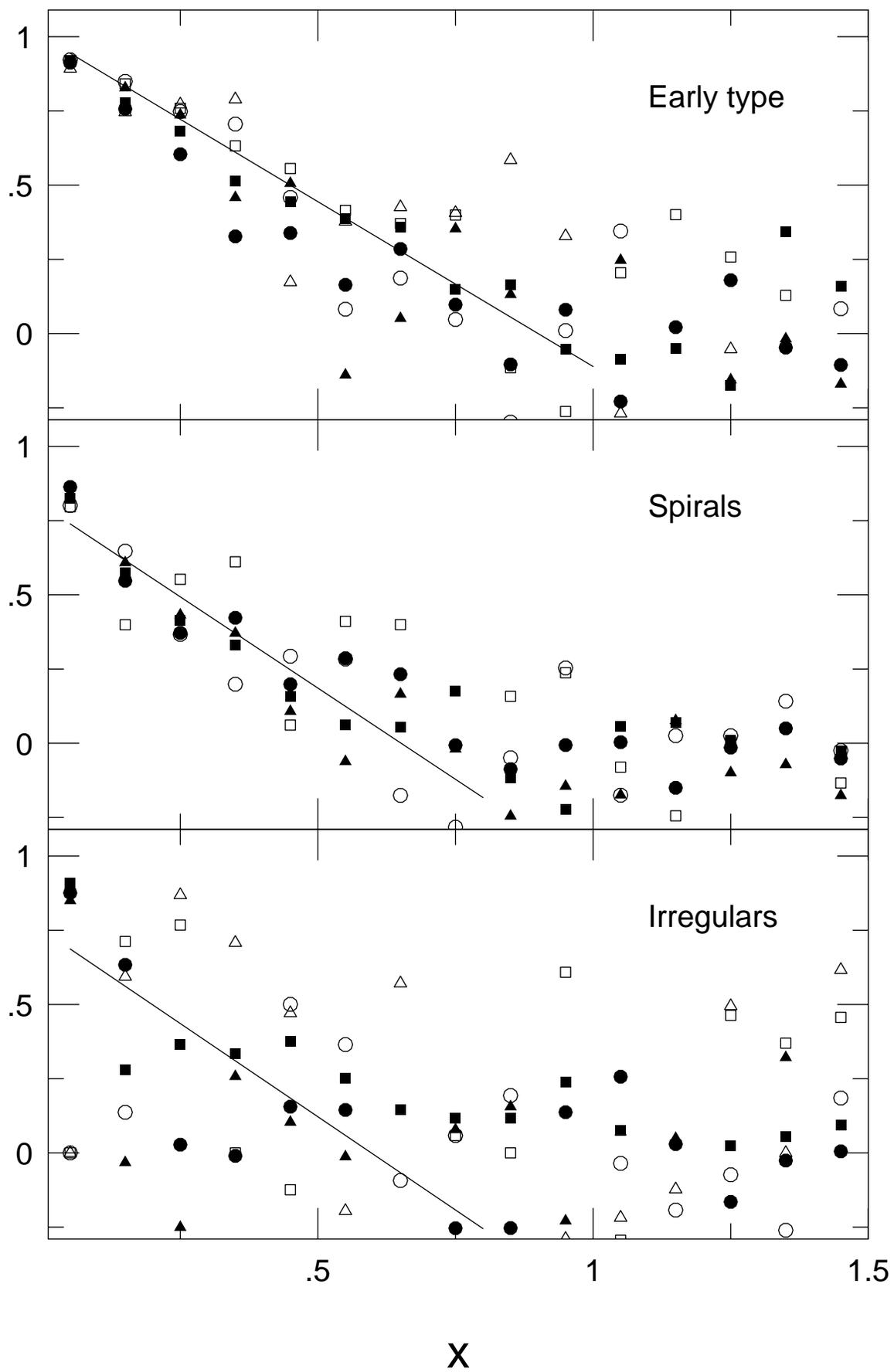

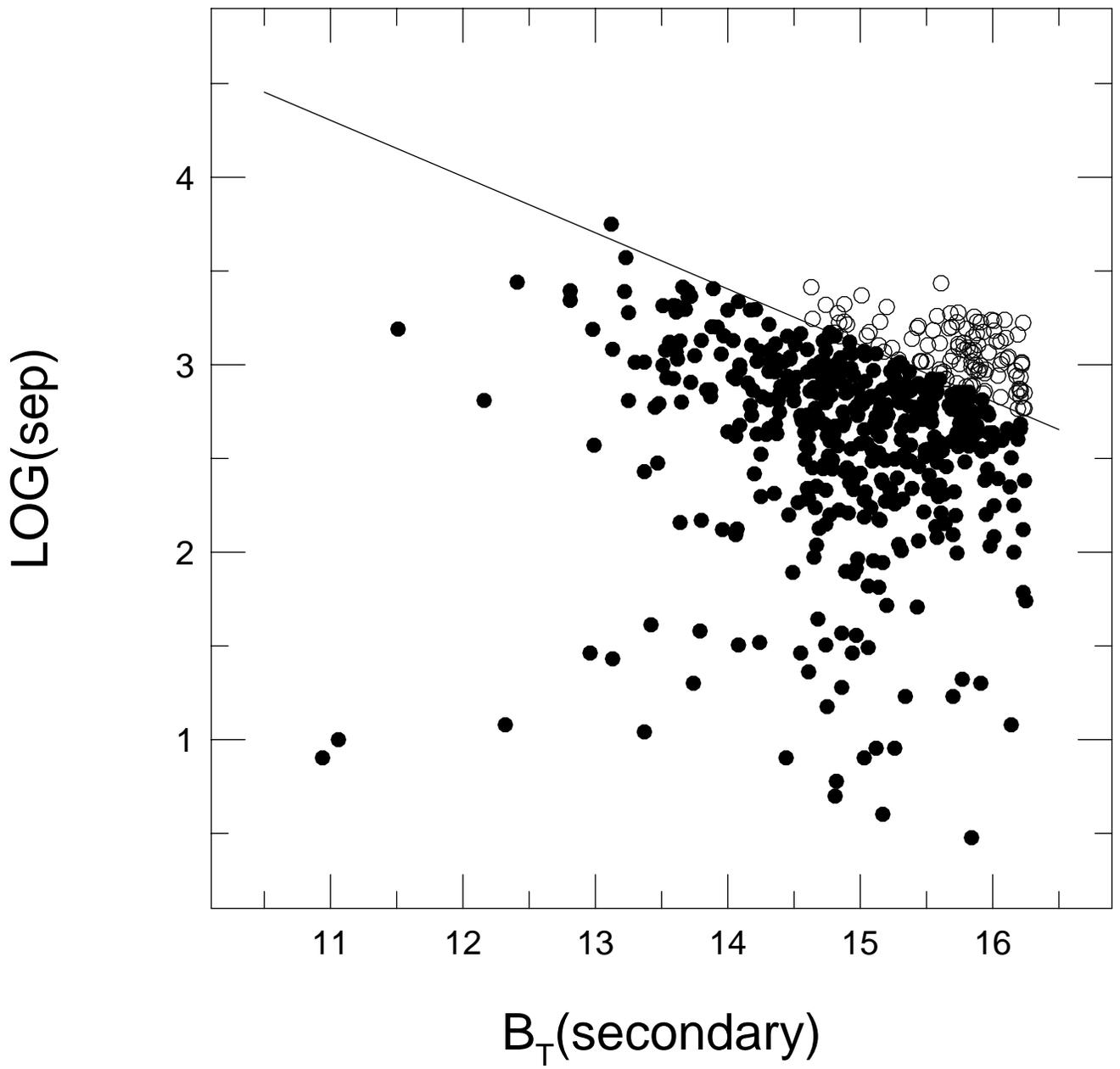

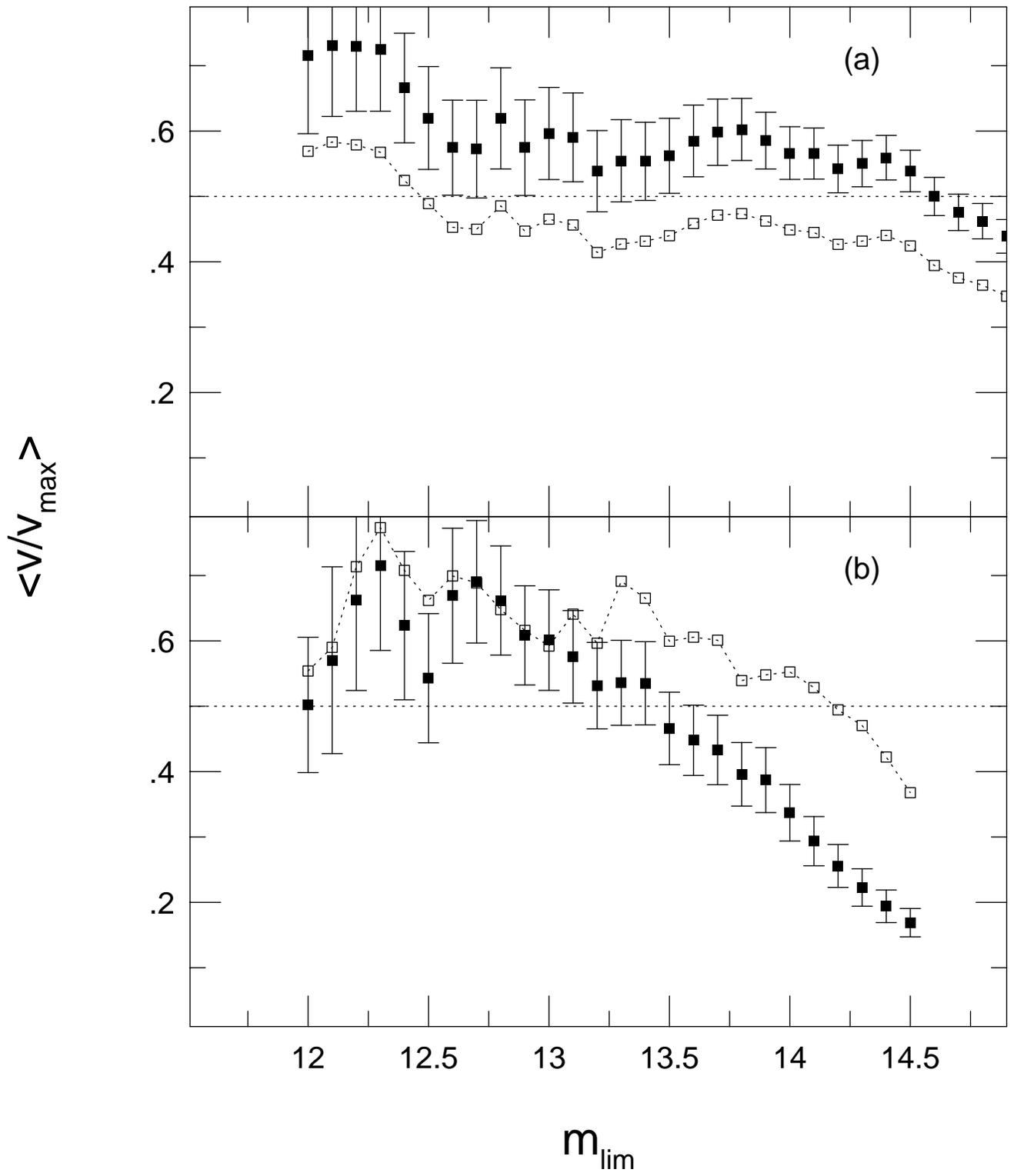

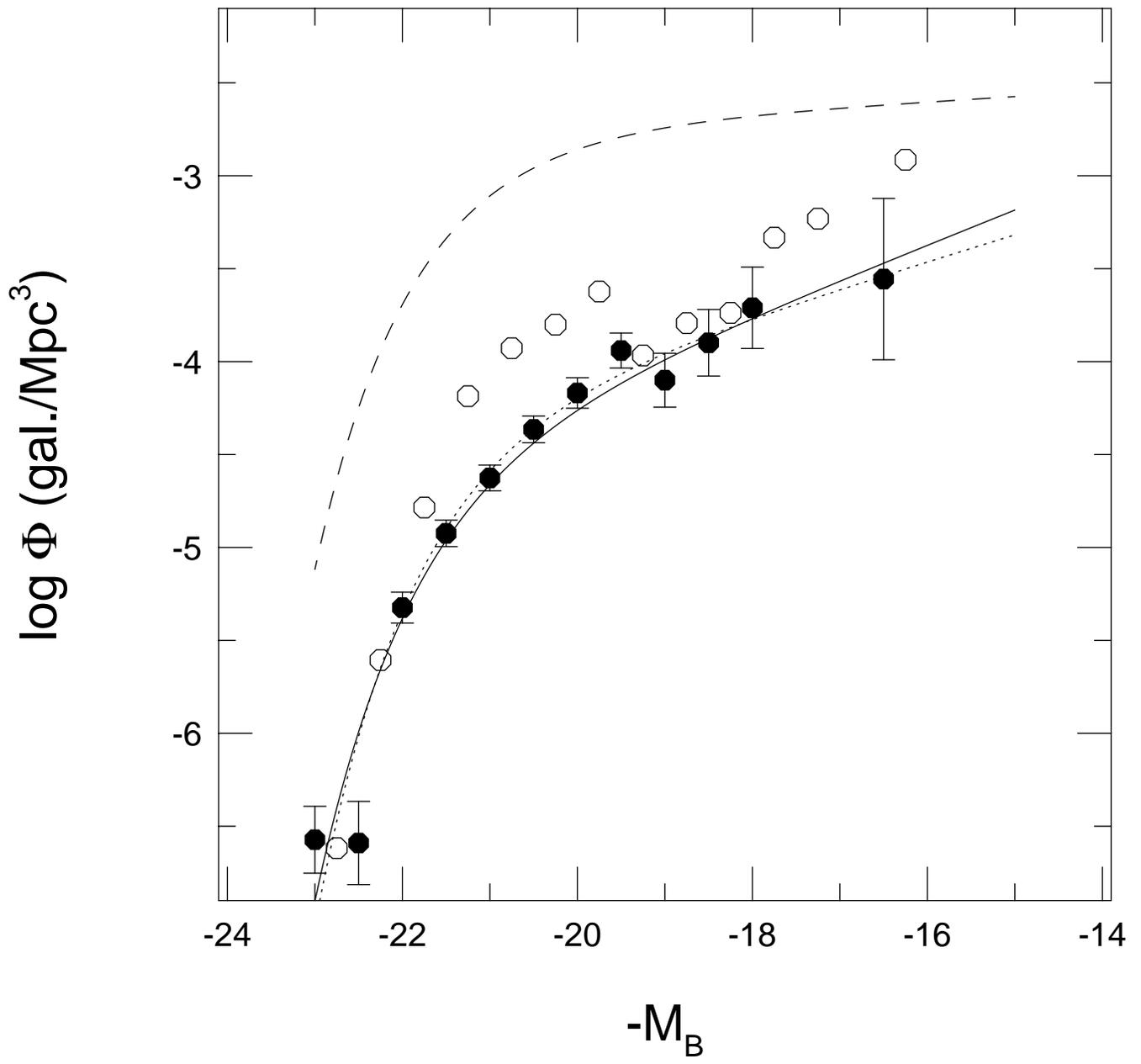